\documentclass[journal, onecolumn]{IEEEtran}

\usepackage[utf8]{inputenc}
\usepackage{amsmath}

\usepackage[T1]{fontenc} 

\usepackage{hyperref} 

\usepackage{mathtools}
\usepackage{graphicx}
\usepackage{tabularx}
\usepackage{subfigure}
\usepackage[table]{xcolor}
\graphicspath{{images/}}
\usepackage{dblfloatfix}
\usepackage{fixltx2e}
\usepackage{multirow}
\usepackage[justification=centering]{caption}

\usepackage{tabularx}					
\usepackage{array}						
\usepackage{booktabs}					

\usepackage{multirow}
\usepackage[table]{xcolor}

\usepackage{amsmath}				
\usepackage{amssymb} 				
\usepackage{mathrsfs} 				
\usepackage{amsthm}				
\usepackage{mathtools}			

\DeclareMathOperator{\oH}{H}			

\newcommand{\sS}{\mathscr{S}}		
\newcommand{\sX}{\mathscr{X}} 		

\theoremstyle{plain}

\theoremstyle{plain}

\theoremstyle{plain}

\theoremstyle{definition}
\newtheorem{definition}{Definition}[section]

\begin{document}
\title{A hybrid analysis of LBSN data to early detect anomalies in crowd dynamics}
\author{Rebeca P. Díaz-Redondo, Carlos Garcia-Rubio, Ana Fernández Vilas, Celeste Campo, Alicia Rodriguez-Carrion
\thanks{Rebeca P. Díaz-Redondo (rebeca@det.uvigo.es) and Ana Fernández Vilas (avilas@det.uvigo.es) are with atlanTTic, Universidade de Vigo; Vigo, 36310, Spain.}
\thanks{Carlos Garcia-Rubio (cgr@it.uc3m.es), Celeste Campo (celeste@it.uc3m.es) and Alicia Rodriguez-Carrion (arcarrio@it.uc3m.es) are with Universidad Carlos III de Madrid, Legan\'es (Madrid), Spain}
}


\maketitle

\begin{abstract}

\noindent Undoubtedly, Location-based Social Networks (LBSNs) provide an interesting source of geo-located data that we have previously used to obtain patterns of the dynamics of crowds throughout urban areas. According to our previous results, activity in LBSNs reflects the real activity in the city. Therefore, unexpected behaviors in the social media activity are a trustful evidence of unexpected changes of the activity in the city. In this paper we introduce a hybrid solution to early detect these changes based on applying a combination of two approaches, the use of entropy analysis and clustering techniques, on the data gathered from LBSNs. In particular, we have performed our experiments over a data set collected from Instagram for seven months in New York City, obtaining promising results.

\end{abstract}



\section{Introduction}
\label{sec:introduction}

The uninterrupted growth in both number of users and activity of Online Social Networks (OSNs) can be attributed to the parallel increase of the smartphone penetration rate. These mobile devices allow a quick interaction with OSNs and make it easy for subscribers to share their ideas, thoughts, photos, messages, etc. All these posts automatically include the subscriber's location when using GPS-enabled devices, especially when interacting with Location-based Social Networks (LBSNs), i.e. location-centred OSNs. These networks focus their activity on sharing experiences at the right place and time they are happening (Foursquare, Twitter, Instagram, etc.).

Consequently, LBSNs provide a very attractive source of geo-located data that, in the smart city field, may be an interesting alternative to the traditional video sources to monitor human activity in urban areas. On the one hand, the infrastructure costs are really low. Instead of having complex video-surveillance networks, which need to be deployed and maintained, citizens are the ones in charge of buying, maintaining and connecting their mobile devices. Besides, due to the ubiquity of the LBSNs, the area under analysis may be easily changed without any extra investment. On the other hand, traditional video-surveillance systems need to be monitored by human staff, who may be supported by complex algorithms for video frames analysis. On the contrary, data from LBSNs would be automatically gathered and analysed (24h-7d) using lighter techniques, like the one we proposed in our previous work~\cite{dominguez2017sensing}. In this former approach we observed that the activity in LBSNs gives trustful information about the usual dynamics of crowds in the city, that can be represented throughout a set of behavioral patterns (one per weekday and time slot of 30 minutes in our proposal). So, the comparison of social media data obtained on-the-fly with these expected behavioral patterns allows detecting unusual activity in urban areas, which may trigger a deeper analysis. Therefore, this approach consumes less computational resources than a video-based techniques, being an appropriate alternative or supplement to the latter.

With the objective of providing a suitable mechanism to early detect unexpected crowds behavior in urban areas, we propose a significant improvement to our previous work, by using a different approach: we select a reduced number of representative points of the area to analyze the entropy of their sequence of locations. Under normal conditions, the behavior of the city (i.e. these representatives) is similar every day, so their entropy values should be also quite similar. Therefore, when something unusual happens, the representatives locations will vary and, consequently, their entropy values will also suffer considerable variations. The technical aspects that have to be faced for this new detection methodology are the following: (i) selecting the city representatives in such a way that their entropy can be analyzed in parallel (to reduce the complexity), but enough to not lose relevant information; (ii) selecting the appropriate sampling frequency, i.e time between each entropy analysis that cannot be too high to avoid unnecessary computations, but not too low to trigger the alert, if needed, as soon as possible; (iii) selecting the appropriate geographic granularity to monitor the movement of the city representatives without loosing relevant information, and (iv) selecting a suitable entropy estimator and an entropy window for the analysis.

This new approach provides a quick mechanism to detect anomalies in the city, but it does not allow to know some important details, like the specific location of the detected ouliers and why these behaviors are considered as anomalies. However, these questions are already answered by our previous methodology \cite{dominguez2017sensing}, which may be triggered when needed. That is, the approach introduced in this paper provides a quick and global analysis, that requires less computational load, whereas our previous approach, with higher complexity, provides a detail analysis only when required. 

The remainder of this paper is organized as follows. Section~\ref{sect:related} provides an overview of other proposals in the crowd and events detection field, especially those which are based on social media data. After summarizing the main techniques used in our approach (clustering and entropy estimators) in Section~\ref{sect:background}, the reasons behind our decision of using Instagram as data source are explained in Section~\ref{sect:dataset}. Our methodology is detailed in Section~\ref{sect:methodology}, whereas its application to the data set and the obtained results are discussed in Section~\ref{sect:results}. Finally, Section~\ref{sec:conclusion} is devoted to conclusions and future work.

\section{Related work}
\label{sect:related}

Early detection of unusual crowds in urban areas is a challenge that has been tackled from different perspectives and using different technologies. Perhaps the most popular are those approaches based on the analysis of data gathered from video-surveillance systems. Smalls groups of pedestrians who are walking together can be detected in \cite{ge2012vision}, motion patterns of large groups are analysed in \cite{wu2014thermal} by using infra-red video data, whereas in \cite{santoro2010crowd} video-data is processed to detect and track crowds. Other proposals are specially focused on the detection of unexpected events using video-data, like in \cite{hamid2005detection} or \cite{xu2016detecting}.

However, the analysis of data gathered from LBSNs has become an interesting option. Proactive LBSNs users can be seen as sensors, constantly providing high volume of data of different nature (text, images, temperature, location, etc.) from the same device (mobile devices). Humans as sensors constitute an alternative or supplement to the costly sensing systems which are traditionally deployed all around urban areas. Additionally, and due to the ubiquity of social media, these applications may be easily used in new areas, without the installation and maintenance costs of dedicated video and/or sensor networks. This advantages have led to the use of this new data source for crowd analysis as well. 

LBSNs provide two kind of valuable information: the location and trajectory of the user posting information, and the content of the posts (text, images, etc.). Regarding the study of the moving patterns that can be obtained from user devices, previous research has been centred in obtaining the most probable next location~\cite{CHEN2015156} and in profiling users based on their mobility patterns~\cite{Trasarti2017350}. Within this area, it is interesting to mention the research work in \cite{shelton2015social}, where the authors explored geotagged Twitter data to analyse everyday activity spaces of different groups of citizens in Louisville (Kentucky, USA) to observe the flow among different neighbourhoods in the city. It is necessary to note that Louisville is, as other American cities, still marked by the legacies of racial segregation that caused discriminatory housing policies that directly affected the neighbourhoods in two different areas in the city: West End and East End. The authors divided the population into these two groups of Twitter users and analysed the relative patterns of mobility of each group. This approach entails the active participation of the users, who must be identified as West End users or East End users. Once this identification is done, the analysis focused on these two groups and their movements within two specific locations: West End area and East End area. 
The analysis of geotagged tweets is also the base of research work in \cite{taubenbock2018poor}. The objective is analysing if the the economic urban divide (usual in any city) is reflected in the digital sphere of cities. This hypothesis was finally confirmed by their study, which shows different behavioural patterns in social media in different neighbourhoods.

Other approach that analyses movement of individuals (users) is explained in \cite{cao2015scalable}, where the flows are studied according to three dimensions: time, location and demographic characteristics (gender, ethnics, age group, occupation and race). They focus on obtaining the space-time trajectories to analyse the users movements throughout North America by taking information from the geotagged posts that they share in Twitter. This research work needs to obtain a large group of users that participate in the experiment by giving their Twitter user-id as well as other demographic data. The trajectories are visualized over different maps (North America, Eastern area of the country, for instance) to see the flows of individuals. 

On the other hand, the analysis of the content of the posts (mainly text) is a challenge that has been tackled from different perspectives. The authors of \cite{arin2018twec} propose a web service able to clustering tweets based on their semantic similarity with the aim of exploring datasets obtained from social media. Being more specific, the analysis of the text published in posts has been repeatedly used for event detection in the literature \cite{weiler2016evaluation,Zhou:2014}. However, and apart from this approach, the location information linked to the shared posts has also brought several proposals to detect crowds and events in urban areas. In \cite{ferrari2011extracting}, most visited locations are detected applying the EM Algorithm to the location of tweets in intervals of two hours. This popular places are associated to a ZIP code. Those ZIP codes are processed using Latent Dirichlet Allocation (LDA) to find patterns in the movements of the crowds and track events with a strong relation with the city. LDA is also applied in \cite{chae2012spatiotemporal}, in this case to the text content of the tweets, in order to find popular topics. Then an abnormality estimation is calculated using Seasonal Trend Decomposition based on Loess smoothing (STL), in an iterative process which requires expert human supervision. \cite{watanabe2011jasmine} also takes advantage of textual content, since it is first used to assign a location to non-geolocated tweets and, after obtaining the popular locations using Geohash, it is used again to decide if the place is popular due to an unusual event or not.

Local events are also the focus in \cite{walther2013geo}, with an approach which constructs clusters of tweets according to the number of tweets in a given area (density). Then, these clusters are scored according to different criteria: textual content, number of users, number of tweets, etc. In \cite{ranneries2016wisdom} posts from both Twitter and Instagram are clustered according to their hashtags. After that, the density-based clustering algorithm DBSCAN is applied twice to these clusters in order to associate a single place to each cluster. A different clustering approach is presented in \cite{lee2010measuring}, where k-means is used to group the geolocated tweets and define Regions of Interest (RoI). Over these regions, the number of tweets is analysed in order to detect outliers. The objective is to develop a geo-social event detection to monitor crowd behaviors and local events. The approach introduced in \cite{lee2012mining} tries to infer spatio-temporal information about the events mentioned in the shared tweets by applying text mining techniques (a density-based online clustering method), with the aim of detecting events in urban areas. In this approach, the location of a tweeted event is obtained by the text analysis, when the geo-tagged information linked to the tweets are not consistent with the text. In \cite{arcaini2016user}, the authors face a different problem: to locate events related to specific themes or aspects that are selected by the user. Public messages posted in LBSNs are the source of information which are gathered (by crawling social networks) and later filtered according to the users' criteria (soccer world cup or floods, for instance). Finally, those messages related to the user's criteria are analysed according to both location and publication time. This exploration is based on the Spatio-Temporal DBSCAN (ST-DBSCAN) clustering algorithm. As a result, it is possible to locate the messages over a local or global map and also to discover periodic and aperiodic events. 
\cite{hasan2017survey} offers a survey on event detection in Twitter and, finally, the authors in \cite{wan2019minor} establishes the base to detect minor probability events in massive data by applying Message Importance Measure (MIM), obtaining promising results in simulated frameworks.

Although machine learning has been the preferred approach in the literature, both supervised and unsupervised techniques, to discover interesting patterns, some proposals combine machine learning with entropy measures over hashtags or words in the posts' content, e.g. \cite{petrovic2010streaming} before applying the analysis or even to filter the obtained results afterwards \cite{xie2016topicsketch}.  Also entropy measures have been applied to the analysis of the structure of social networks, e.g. \cite{nikolaev2015efficient} defines entropy-based centrality to provide  a community detection being more flexible and efficient than community discovery with traditional centrality measures;\cite{yang2017hypergraph} describes a hyper graph partitioner which incorporates modularity based on information entropy to achieve a more scalable solution over social structures. Closer to the mobility scenario in this paper, \cite{yuan2015hotspot} proposes an entropy-based solution to forward data along opportunistic social networks. The new routing metric Hotent (HOTspot-ENTropy) exploits hotspot entropy to design an utility function which computes the centrality of  the nodes as the relative entropy between the public hotspots and the personal hotspots; and similarity between two nodes as the inverse symmetrised entropy of personal hotspots. Also, entropy measurement has been proposed in ~\cite{rodriguez2015entropy}, in the context user mobility patterns from GSM traces, to predict users’ next location an so  to anticipate their future context. \cite{electronics} proposes a new entropy-based methodology for early detection of anomalies in urban areas that exploits the location data of the posts published on LBSNs. The proposal uses just one centroid as the single geographic point summarizing the state of the city, the location of which is tracked to detect changes in its entropy evolution. Finally,  entropy has been used in the context of social sharing in a pervasive Web \cite{CudrMauroux2008} to face the problem of metadata scarcity by a system proposal in which information entropy—in terms of missing metadata—is gradually alleviated through decentralised instance and schema matching.

Finally, our previous work \cite{dominguez2017sensing} stands out because of the following reasons: (i) it does not need the high complexity of video-based approaches; (ii) it does not need specific actions of the citizens (installation of any specific application in their mobile devices and/or include specific tags in their posts), they only need to act as they usually act and publish their posts in LBSN as they normally do; (iii) it establishes patterns of crowds dynamics throughout the city by using an improved density-based clustering algorithm; and (iv) it is able to detect not only unusually big crowds, but also unusually small crowds, crowds in areas that often have much less activity, and the absence of crowds in areas that often have higher activity. 
With the aim of reducing the computational load of our previous methodology, we propose in this paper to substantially improve the detection of anomalies using a hybrid approach that combines entropy analysis and clustering techniques.

\section{Background}
\label{sect:background}

Our approach tries to detect unusual activity in social media, so in this section we summarize some of the techniques that we have used for this purpose: (i) some of the mainly known clustering techniques (in Section~\ref{sect:backgroundClustering}), which was used to organize public posts according to the GPS location from where were published; and (ii) the main concepts behind entropy (in Section~\ref{sect:entropy}), which was used for the anomaly detection.

\subsection{Clustering techniques}
\label{sect:backgroundClustering}

Clustering is ``the process of partitioning a set of data objects into subsets or clusters such that objects in a cluster are similar to one another, yet dissimilar to objects in other clusters''~\cite{clustering}. Clustering methods are generally classified in four groups: partitioning approaches (where the number of clusters is pre-assigned), grid-based (where the object space is divided into a pre-assigned number of cells), hierarchical (where the data is organised in multiple levels) or density-based (where density notion is considered). We have applied two different strategies, a partitioning and a density-based algorithm, which will be used for different purposes in our approach. 

K-means algorithm is the most widely known partitioning clustering method because of its simplicity: it simply partitions a data set into $\{C_1, C_2, \ldots, C_K\}$ distinct, non-overlapping clusters. To perform this technique, parameter $K$ (the desired number of clusters) is firstly specified. Then, the algorithm will assign a cluster $C_{j}$ to any data point $i$ such as the total within-cluster variation $W(C_k)$, summed over all $K$ clusters, is as small as possible, i.e. the problem to be solved is the following one:
$$
\underset{C_1, C_2, \ldots, C_K}{\text{minimize }} \{ \sum_{k=1}^{K} W(C_k)\}
$$
where $W(C_k)$ measures the amount by which the observations within a cluster differ from each other.

Density-based clustering algorithms organize the data in regions (clusters) where the elements are dense and which are separated by areas of low element density, which are considered as noise. These algorithms are able to (i) discover clusters of arbitrary shapes, (ii) handle sparse regions (which are considered as noise regions) and (iii) work without knowing the number of clusters in advance. Among the different proposals in the literature \cite{dbscancomparative}, we have selected DBSCAN~\cite{dbscan}(Density-Based Spatial Clustering of Applications with Noise), since it does not add any extra-functionality and so extra-computational load.

DBSCAN needs two parameters to define the density of the clusters: the minimum number of elements ($minPoints$) that must be located within a given distance $\epsilon$ from an element in order to start forming a cluster. In fact, DBSCAN defines an element as a noise point if it is not enough close to other elements, otherwise DBSCAN assigns the element to a particular cluster. For this, DBSCAN determines the local density at each element by using the previous two parameters: reachability parameter (distance value $\epsilon$) and the minimum number of elements ($minPoints$). An element or point $p$ which meets the minimum density criterion, i.e. there are at least $minPoints$ located within a distance $\epsilon$ from $p$ is considered a core point and defines an $\epsilon-neighbourhood$. Any element or point $q$ within the $\epsilon-neighbourhood$ is considered as directly reachable from $p$. Any element or point $q$ is reachable or connected by density from $p$ if there is a path of elements or points that connect both elements through chains of $\epsilon-neighbourhoods$. Therefore, a core point forms a cluster together with all elements or points that are reachable from it. As aforementioned, those points that are not assigned to any cluster are considered as noise.


Once a clustering technique is applied, the quality of the obtained clusters should be assessed. There are several 
quality measures that can be used for this purpose, although silhouette~\cite{rousseeuw1987silhouettes} is perhaps the most popular. The silhouette value is a measure of cohesion (similarity among the data points in the same cluster) and separation (dissimilarity among data points classified in different clusters). The silhouette value for a data point $i$ is obtained as follows:

  $$ s(i) = \begin{cases}
    1-a(i)/b(i)    & \quad \text{if } a(i) < b(i)\\
    0 & \quad \text{if } a(i) = b(i)\\
    b(i)/a(i)-1 & \quad \text{if } a(i) > b(i)\\
  \end{cases}$$

Where $a(i)$ is the average dissimilarity of $i$ with respect to the other data points in the same cluster and $b(i)$ is the lowest average dissimilarity of $i$ with respect to any other cluster. The silhouette value ranges from $-1$ and $1$: a high value indicates the data point has been correctly classified, whereas low or negative values show the clustering should be improved. 

\subsection{Entropy concept and estimators}
\label{sect:entropy}

This section provides an introduction to the concept of entropy and its practical interpretation. A wider review on this topic can be found in~\cite{gao2008,cover06B}. Starting with the basics, next we introduce the definition of what is known as \textbf{Shannon entropy} in the information theory domain, introduced by Claude E. Shannon~\cite{shannon1948}.

\begin{definition}
	Let $X$ be a discrete random variable taking values on an alphabet $\sX$, being $|\sX|$ the cardinality of the alphabet, and with PMF $\Pr\left(X=x\right)=p\left(x\right), \forall x\in\sX$. Then, the entropy can explicitly be written as:
	\begin{equation}
	\oH_S\left(X\right)=-\sum_{x\in\sX}p(x)\log_2{p\left(x\right)}
	\label{eqn:shannonEntropy}
	\end{equation}
\noindent where the base 2 logarithm denotes that the resulting entropy value is measured in bits.
\end{definition}

Paying attention to the practical meaning of entropy, $\oH\left(X\right)$ measures the expected ``surprise'' or uncertainty enclosed by the random variable $X$.

With the initial concept of entropy it can be hard to measure the real randomness of a sequence of events. Besides the probability of each symbol, there exists information regarding the temporal correlations among one sample and the previous ones. Therefore, we here present some estimators that help to cope with this problem.

%
%


%

Let $(X_n)$ be an stochastic process, taking values on the alphabet $\sX$, with cardinality $|\sX|$. Let $S=s_{1}s_{2}\ldots \allowbreak s_n\ldots s_{N}$, be the finite sequence of observed outcomes of a realisation of $(X_n)$, of length $N$.  Therefore, the set of different values that $s_n$ can take on, $\sS$, is a subset of $\sX$, $\sS\subseteq\sX$, with cardinality $|\sS|\leq|\sX|$.

The Hartley entropy estimator, $\oH_H\left(X_n\right)$, is defined as the number of different symbols observed in the available sequence, $S$:
\begin{equation}
\hat{\oH}_H\left(X_n\right)=\oH_H\left(S\right)=\log_2{|\sS|}
\label{eqn:hartleyEntropyEstimator}
\end{equation}

For calculating the Shannon entropy estimator, since the probability mass function $p\left(s_j\right)$ is not available, it is approximated by a maximum likelihood estimator based on the observable data. Let $N\left(s_j\right)$ be the number of elements, $s_n$, of the observed sequence, $S$, that are equal to $s$: $N(s)=|\{n\in \{1,2,\ldots,N\}: \allowbreak s_n=s\}|$. Then, the estimator  of $p\left(s_j\right)$, is then calculated as follows: 
\begin{equation}
\hat{p}\left(s\right)=\frac{N\left(s\right)}{N}, \forall s\in\sS
\label{eqn:probabilityMassFunctionEstimator}
\end{equation}

Therefore, the estimator for the Shannon entropy can be expressed as:
\begin{equation}
\hat{\oH}_S\left(X_n\right)=\oH_S\left(S\right)=-\sum_{s\in\sS}\hat{p}\left(s\right)\log_2\hat{p}\left(s\right)
\label{eqn:shannonEntropyEstimator}
\end{equation}

Estimating the entropy rate of a finite sequence is more complex. In fact, the complexity stems from the problem of not having enough samples of the sequence so as to completely capture the probability mass function describing the model underneath. In ~\cite{grassberger1989,kontoyiannis1998} it is presented an alternative entropy rate estimator that avoids the effect of not having enough data, so it is possible to accurately estimate the entropy rate of a sequence. These authors propose an entropy estimator based on block lengths:
\begin{definition}
	The Grassberger entropy rate estimator is expressed as:
	\begin{equation}
		\hat{\oH}_R\left(X_n\right)=\oH_R\left(S\right)=\left(\frac{1}{N}\sum_{i=2}^{N}\frac{\Lambda_i}{\log_2i}\right)^{-1}
	\label{eqn:grassbergerEntropy}
	\end{equation}
\noindent where $\Lambda_i$ is the length of the shortest substring starting at index $i$ of the sequence, $S$, that did not appear in the range $\left[1,i-1\right]$, where $N$ is the length of the whole sequence.
\label{def:grassbergerEntropy}
\end{definition}

Since we need to calculate the entropy at each new data point, we used the following approximation for the probability of a given location, and apply it to the entropy formula for each time interval, $i$.
\begin{equation}
H(i)=-\sum_{l\in\mathscr{L}}p(l,i)\log_2 p(l,i); p(l,i)=\frac{N_{l,i}}{i}, 0\leq i\leq n
\label{eqn:grassbergerEntropy2}
\end{equation} 
\noindent where $N_{l,i}$ is the number of appearance of location $l$ in the sequence from the beginning up to time interval $i$, and $n$ is the total number of time slots.

\section{Data set}
\label{sect:dataset}

The selection of the data set is important, since it must be data gathered (i) from publicly accessible geo-located posts obtained from LBSNs; (ii) for a large time interval, in order to analyze the evolution of the activity in social media over time and (iii) that allows a comparison of the results of this new approach with our previous work. These are the reasons why we have decided to use the same data set.

When we faced the problem of gathering a data set, we studied three different alternatives: Twitter, Instagram and Forsquare. Twitter was discarded because of the important restrictions of the two available APIs. As a summary, Twitter Search API limits the number of calls both per user and per application, whereas Twitter Streaming API allows accessing only to the 1\% of the published tweets and the kind of sampling is not public. Foursquare posts, on another hand, are always linked to public venues (restaurants, museums, etc.) since its objective is sharing opinions about them. Therefore, Foursquare posts are biased by the venues locations. Finally, Instagram turned out to be the best choice because of the following reasons: (a) it allowed gathering posts published any moment in the past; (b) it imposed less calls restrictions and (c) the data was not biased by specific locations. 

One of the limitations of using Instagram or any other social network to detect anomalous events is that it is very sensitive to changes in API policies. At the time we carried out the New York data capture campaign (end of 2015 and beginning of 2016), the Instagram API media search endpoint allowed recovering data from any moment in the past, between two timestamps\footnote{The API at the time of our data capture campaign at the Internet Archive is available at {\scriptsize http://web.archive.org/web/20150531210319/https://instagram.com/developer/endpoints/media/}}, so our data set could be re-extracted later by anyone (for example, to reproduce our study). This is not possible anymore. Besides, since June 2016, for privacy reasons, the media/search endpoint in Sandbox mode was limited to return just the media you uploaded from that location. To have access to the public content published by others, you need to submit your application to Instagram for approval for the Live mode. This was not required when we made the New York capture campaign. However, and although the policy to gather data from Instagram has substantially changed, it does not change the results obtained in this research work, since the methodology can be applied to any available LBSN data.

For gathering the data, we firstly set the area under study to the circular area centered at Times Square (40.756667 N, 73.986389 W) with a radius of 5 Km. (the maximum allowed by the API). Secondly, we set a wide extraction period, from 23rd August 2015 to 28th February 2016 (190 days) which covered special days (Table\ref{table:specialEvents}), when the city is traditionally more crowded (like Christmas time) or when the city was less crowded (like the weekend when Storm Jonas hit the United States), and also days which are considered normal, when no special events or phenomena are expected to happen. During this period, we collected $4,335,880$ posts, which are distributed as shown in Table~\ref{table:postsDistribution}.(a) and Table~\ref{table:postsDistribution}.(b). It should be remarked that the set of {\em special} days, where selected knowing that events like {\em Columbus Day}, {\em Christmas' Eve} or {\em Veterans Day} clearly affect the normal crowd dynamics.

\begin{table}[!htbp]
\centering
\begin{tabular} {cc}
  \hline
 Date & Event\\ 
  \hline
2015/09/07	&	Labor Day \\
2015/10/12	&	Columbus Day \\
2015/10/31	&	Halloween \\
2015/11/11	&	Veterans Day \\
2015/11/26	&	Thanksgiving Day \\
2015/12/24	&	Christmas' Eve \\
2015/12/25	&	Christmas \\
2015/12/31	&	New Year's Eve \\
2016/01/01	&	New Year \\
2016/01/21-24	&	Jonas' Storm 

\end{tabular}
\caption{Special events considered}
\label{table:specialEvents}
\end{table}

\begin{table}[!htbp]
\begin{tabular} {cc}

\parbox{.45\linewidth}{
\centering
\begin{tabular}{ccc}
  \hline
 Month & Total & Average\\ 
  \hline
 2015-08 & 162,602 & 20,325.25 \\ 
 2015-09 & 688,405 & 22,946.83 \\ 
 2015-10 & 720,607 & 23,245.39 \\ 
 2015-11 & 666,256 & 22,208.53 \\ 
 2015-12 & 697,927 & 22,513.77 \\ 
 2016-01 & 723,445 & 23,336.94 \\ 
 2016-02 & 676,638 & 24,165.64 \\ 
 \hline
 & 4,335,880 & 22,677.48 \\

   \hline
\end{tabular}
}

&

\hfill
\parbox{.55\linewidth}{
\centering
\begin{tabular}{ccc}
  \hline
 Day & Total & Average \\ 
  \hline
  Mon & 615,309 & 22,789.22 \\
  Tue & 609,233 & 22,564.19 \\ 
  Wed & 610,320 & 22,604.44 \\ 
  Thu & 629,413 & 23,311.59 \\    
  Fri & 621,170 & 23,006.30 \\ 
  Sat & 622,923 & 23,071.22 \\ 
  Sun & 627,512 & 23,241.19 \\  
\hline
 & 4,335,880 & 22,941.16 \\
   \hline
\end{tabular}
}

\\
& \\
(a) Per month  & (b) Per day of the week \\
& \\
\end{tabular}

\caption{Posts distribution}
\label{table:postsDistribution}
\end{table}

\section{Methodology}
\label{sect:methodology}

\begin{figure}
  \centering
   \includegraphics[width=\columnwidth]{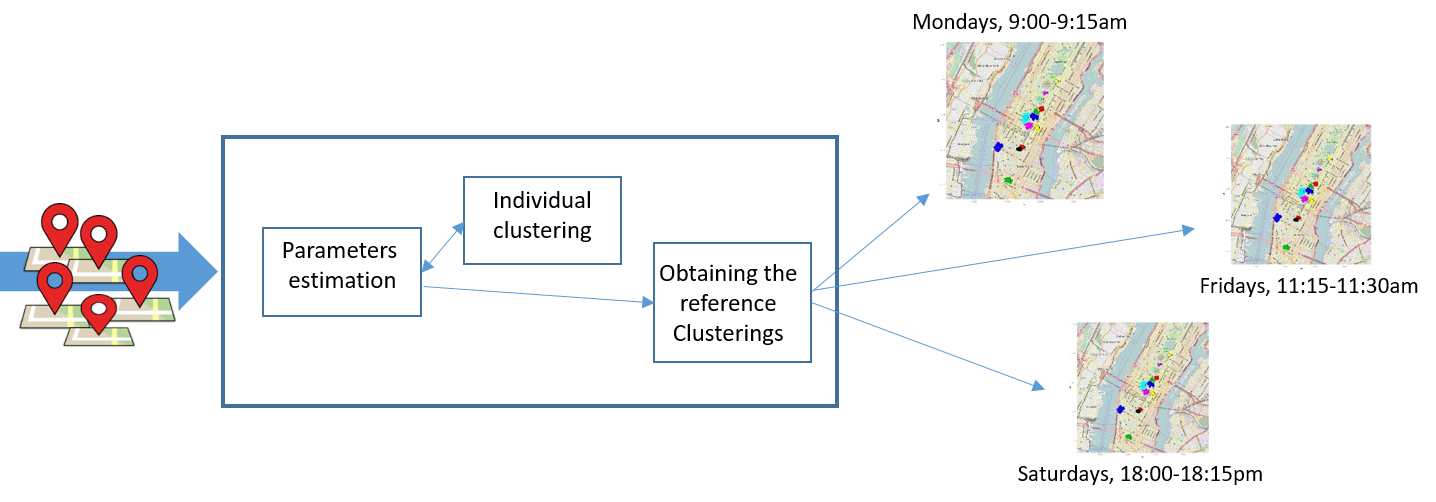}
	\caption{First Stage: Obtaining behavioural patterns}
\label{fig:TrainingStage}
\end{figure}

In our previous work we have corroborated that sensing the activity in LBSNs is a suitable mechanism to obtain behavioral patterns of the crowd dynamics in an urban area (First Stage in Fig. \ref{fig:TrainingStage}). Based on an improved density-based clustering (DBSCAN), we have obtained different {\em reference clusters} or behavioral patterns that show the usual social media activity distribution throughout the city. Each reference cluster is obtained for a specific day of the week and for a specific time interval, for instance Mondays from 9:00am to 9:15am, and represents the usual activity for any Monday at the same time interval. Of course, this process may be adapted for different criteria of analysis: different time intervals, or working days and no-working days, for instance.

\begin{figure}
  \centering
   \includegraphics[width=\columnwidth]{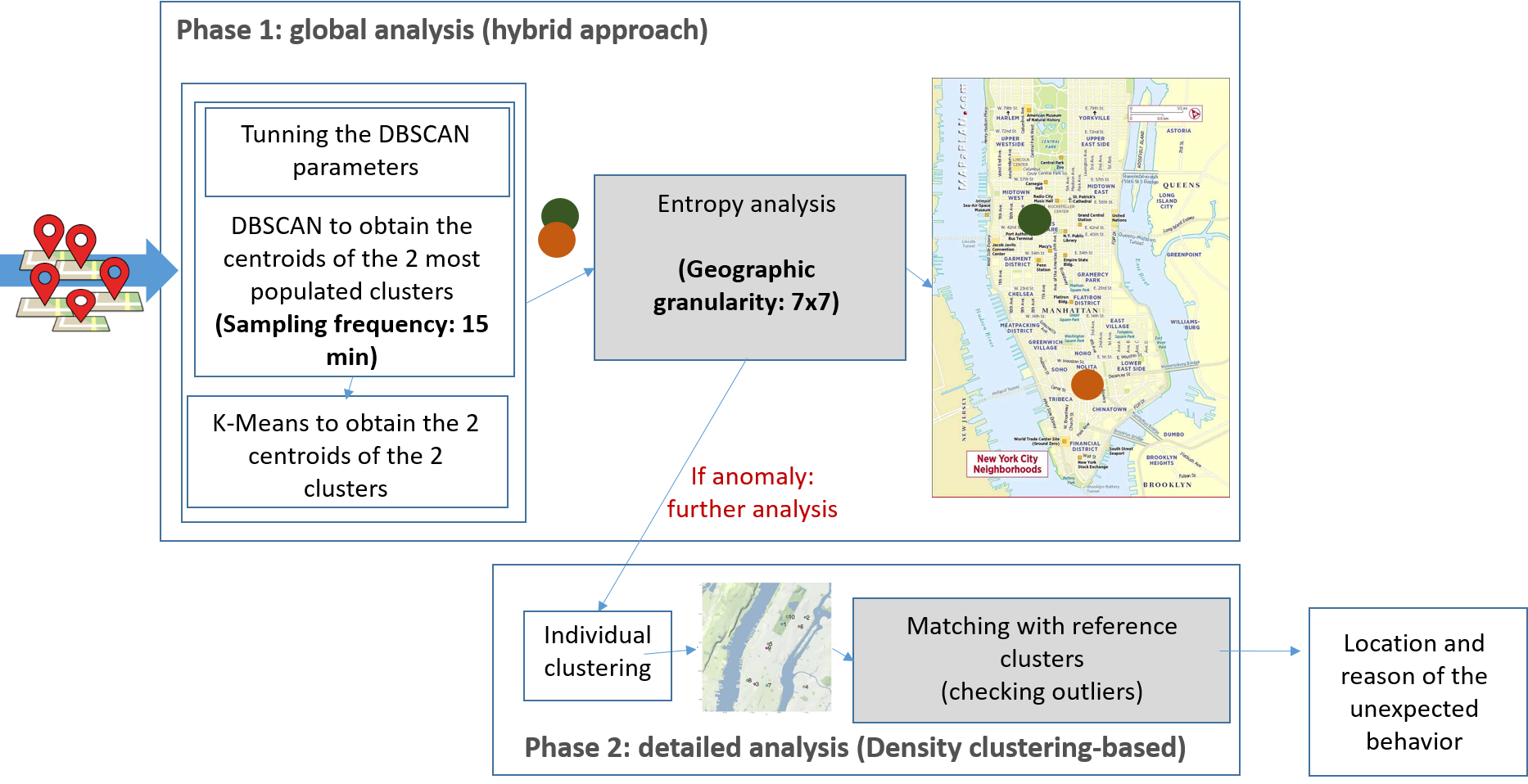}
	\caption{Second Stage: Anomaly Detection}
\label{fig:DetectionStage}
\end{figure}

In this paper, we have not focused on the First Stage (Obtaining behavioural patterns), but in the Anomaly Detection (Second Stage), changing substantially our previous approach by combining two techniques: entropy analysis and clustering techniques. As Fig.\ref{fig:DetectionStage} depicts, the Anomaly Detection Stage is composed of two phases. Phase 1, which provides a global analysis, only analyses a reduced number of data points, which were coined as {\em representative points} and it does not need any behavioral pattern or reference clustering to detect if something unusual is happening in the global LBSN activity. Phase 2, which provides a detailed analysis, is triggered when Phase 1 detects any anomaly, and is responsible for realizing a further analysis to detect where the unexpected behaviors are located and why they are considered abnormal (caused by a higher or lower activity than expected according to the reference cluster and/or caused by activity happening in a different location than expected, for instance).

%

For this new Phase 1 to work properly it is essential to make decisions on the following four technical aspects. Firstly, it analyzes the behavior of only a few geo-located points. Thus, these {\em representative points} should be carefully selected. Section~\ref{sec:cityRepresentatives} and Section~\ref{sec:numberCentroids} describe in detail this selection process. 

Secondly, the movement of these {\em representative points} all around the city is monitored along time, so it is needed to select an appropriate sampling frequency to harvest the geo-located posts: not too high, in order to have enough data for the analysis and to avoid unnecessary computations; but not too low, since an early detection is pursued. This procedure is described in Section~\ref{sec:frecuency}.

Thirdly, the representatives' coordinates will be transformed into the symbolic domain to compute the entropy of the sequence later on: each position will be represented by a different symbol. Thus, in order to maintain the cardinality of the alphabet in reasonable orders, the city is split into non-overlapping squares of the same size (a grid). Then, each square or cell is labeled with a symbol. Note that although the Instagram API gives us posts in a circular area, what we want to discretize in a series of cells is not the position of all the posts but the coordinates of the representatives obtained with the DBSCAN algorithm. The position of these representatives  does not necessarily follow a circular distribution, but depends also on the geography of the city and where its points of interest are located. Examining the first two months of the New York trace we have observed that a square area of diameter 5 Km included all the representatives obtained with DBSCAN, so we decided to use a square grid, simpler than a circular one. Then the position of each representative at each temporal interval is identified by the symbol of the cell in this grid that encloses that position. Therefore, in case of having more than one representative points distributed along the cells, the city behavior would be expressed as a $n$-tuple containing the cells where each of the representative points are located at each specific time. Therefore, we need to define the size of these smaller areas, i.e. the spatial granularity of the analysis. Again, not too low, in order to have accurate results and be able to properly locate the potential focus of the problem; but not high, since we do not want to unnecessarily increase the cardinality of the alphabet. This process is detailed in Section~\ref{sec:frecuency}.

Fourthly, we quantify the behavior of the movement of each representative point as the deviation from the expected uncertainty of that representative's movement. People movements have some degree of randomness, as shown in~\cite{gonzalez2008understanding}, and so does the behavior of the resulting crowd. However, big deviations from the expected value of uncertainty can potentially unveil unexpected events. This way, we allow the point movement to have the expected level of randomness, but we aim to capture the times in which that randomness is too different from the expected value. One way to measure the expected uncertainty of a sequence of symbols drawn from an alphabet $\mathscr{L}$, is through the entropy estimators described in Section~\ref{sec:entropywin} and Section~\ref{sec:entropyestimator}.

Finally, we inspect the values of the entropy calculated at each time interval, $i$, and label as potential anomalies those samples with higher entropy differences with respect to the previous value.

\subsection{Procedure to select the city representatives}
\label{sec:cityRepresentatives}

The selection of the geo-located points that will be the city representatives is key. We need to have a reduced number of representative points such as they could be considered independent and, consequently, their entropies could be analyzed in parallel. Taking into account that the urban area under study has a 5 km radius from the center location, we have considered selecting one, two or three points. These selected points are intended to represent the movement of the citizens all around the urban area for a specific time interval or time-slot. 

The first option, taking only one representative point (centroid) was faced in \cite{electronics} with promising results. Although there are several alternatives, like applying an average latitude/longitude method or a centre of minimum distance method, we have used a geographic midpoint method using the Haversine distance, to take into account the points are in the surface of the Earth. In this paper, we approach the other two options, selecting 2 or 3 points as representatives, we have studied three different alternatives:
\begin{itemize}
\item {\bf Opt. 1} Applying the K-Means clustering algorithm, using the Haversine distance for assessing the within-cluster variation, and setting the number of expected clusters to 2 and 3, respectively. The representative points will be the respective centroids of the resulting clusters.
\item {\bf Opt. 2} Applying the DBSCAN clustering algorithm. After having set the parameters $\epsilon$ and $minPoints$, the algorithm infers a set of clusters. Selecting the 2 or 3 most populated clusters as representative clusters and their respective centroids as representative points.
\item {\bf Opt. 3} Applying a hybrid technique which combines the two previous clustering techniques. Firstly, the DBSCAN algorithm is applied as outlined in option 2. Then, the 2 or 3 representative points are used as the input parameters for the K-Means algorithm to fix the initial centers. Finally, the K-Means clustering is applied as outlined in option 1.
\end{itemize}

The selection of the most suitable option should be based on the quality of the resulting clustering as a measure of how independent the clusters are. A low clustering quality entails an unnatural partition of the data set and, consequently, the independence premise is not fulfilled. As it was introduced in Section~\ref{sect:backgroundClustering}, the silhouette is the most popular quality measure and the one we have used to assess the quality of the three aforementioned options. We have proceed as follows: (i) selecting the temporal granularity of the analysis (15 min. and 30 min.); (ii) selecting the day of the week (Mondays, Tuesdays, etc.); (iii) applying one of the three clustering options; and (iv) running the silhouette algorithm. It should be remarked that when talking about selecting a day of the week it does not mean we only do the analysis with the data gathered of only one day. On the contrary, we select one day of the week, Mondays for instance, and we gather the data of every single Monday in the period of analysis ($190$ days) that entails a total of $615,309$ posts according to the information detailed in Table~\ref{table:postsDistribution}.(b). 

\begin{figure}
  \centering
   \includegraphics[width=0.8\columnwidth]{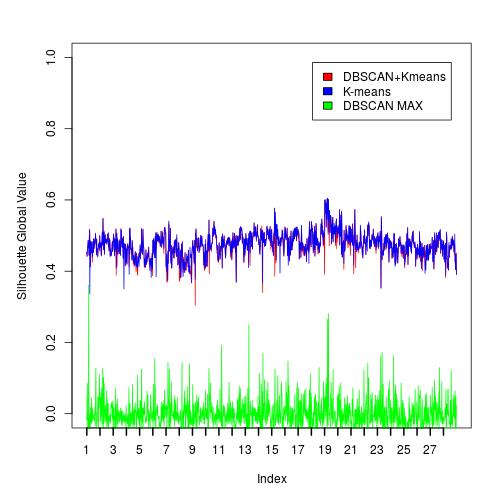}
	\caption{Clustering quality results: Saturdays, 30 min. periods and 2 representatives.}
\label{fig:no3clusters}
\end{figure}

Therefore, the analysis was exhaustively done to analyse the quality of all possible combinations. As an example, Figure~\ref{fig:no3clusters} shows the quality results applying the three alternatives over the data set for Saturdays organised into 30 min. periods. As it is clearly noticeable, the second alternative (DBSCAN only) shows poor quality results, compared to the other two options (first and third). Since this behavior was repetitively observed in all the analysed combinations, we decided to discard the second approach to obtain the representatives.

The decision between the first and the third alternatives, K-means and the hybrid solution respectively, was not only based on the quality results, since both of them offer quite similar measures. Instead, we focused on other desirable characteristics, like reproducibility. The hybrid approach allows to repeat the analysis (same data set and algorithm sequence) and obtain identical results, whereas the K-means alternative does not guarantee so. Therefore, the clustering technique we have decided to use is the hybrid solution: applying the DBSCAN algorithm for identifying the 2 or 3 representative points and, after that, applying K-means to identify the groups.

\subsection{How many representatives are needed?}
\label{sec:numberCentroids}

\begin{figure}[!htbp]
  \centering
	 \includegraphics[width=0.7\columnwidth]{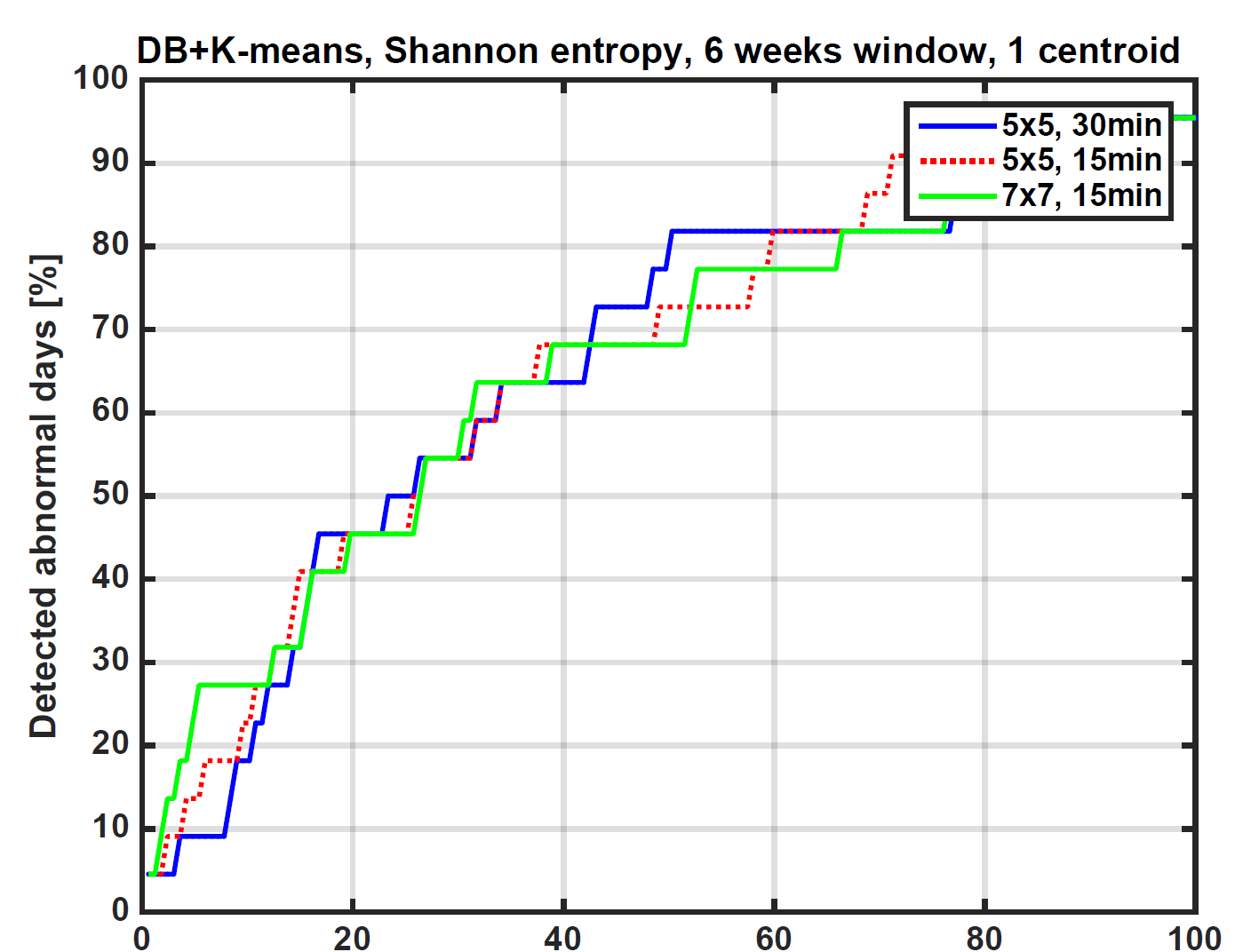}\\
	 \includegraphics[width=0.7\columnwidth]{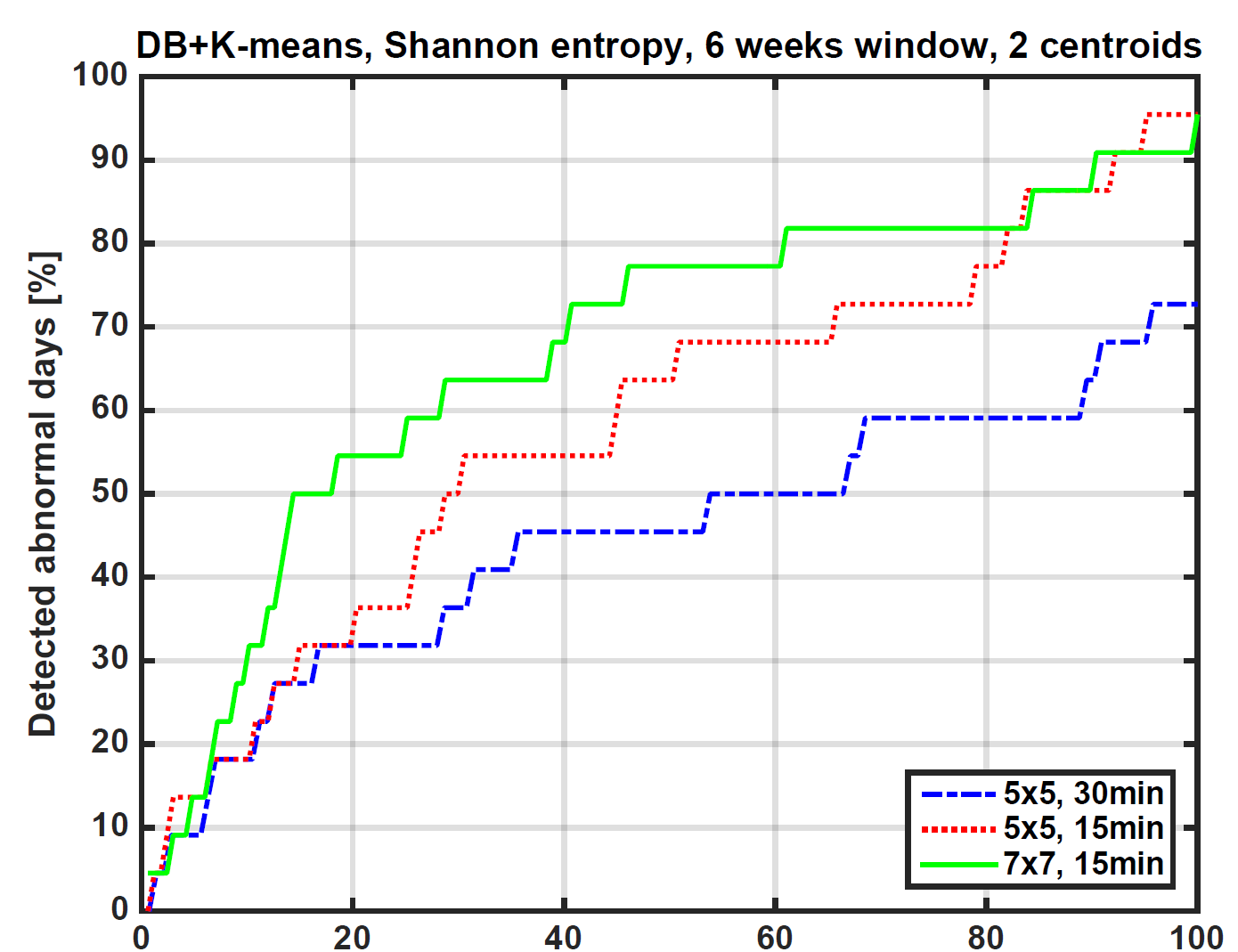}\\
	\caption{One or more than one representatives. Abnormal day detection: percentage of abnormal days detected vs. percentage of days processed.}
\label{fig:abnormalDayDetectionResults}
\end{figure}

Once the criteria for selecting the city representatives has been determined, the next step is deciding how many representatives are appropriate to characterise the city behavior. The first step is deciding if it is better to have only one representative or more than one. The criterion we have applied is to check which option offers better results when detecting unusual days. The process consisted in applying the entropy calculus to detect the already known special days (Table~\ref{table:specialEvents}) for the two alternatives: one or more than one representatives. \figurename~\ref{fig:abnormalDayDetectionResults} shows the results: percentage of abnormal days detected with respect to the number of days processed. These values are obtained if we take the days with the highest entropy value difference between the beginning and the end of the day, and ignoring the first month of data to filter out the initial values which might be misleading. 

If we compare the first plot (one centroid) with the second one (two centroids), we can see that the results for any combination of temporal and spatial granularity performs very similarly for the one centroid case, whereas the performance is quite different in the two centroid case. Comparing the overall percentage of detected abnormal days, the one centroid case reaches around 50\% of detected days when 20\% of days are processed, and between 70 and 80\% when 60\% of days are processed. The two centroid cases outperform these results, thus detecting more abnormal days having processed less of the total days. Therefore, in taking into account that we have obtained promising results in our previous work with only one centroid \cite{electronics}, the results now should be even better.

\begin{figure}[!htbp]
  \centering
   \includegraphics[width=0.7\columnwidth]{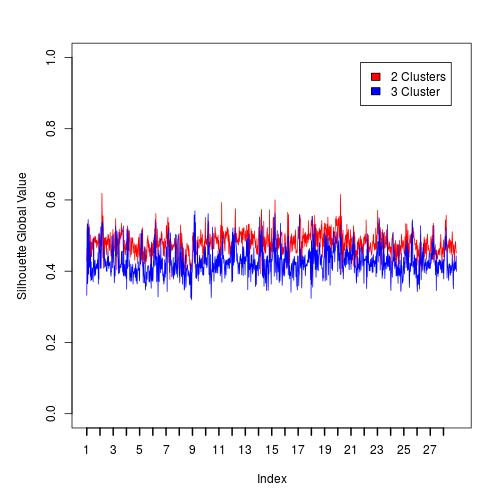}
	\caption{Clustering quality results: hybrid solution, Mondays, 30 min. periods.}
\label{fig:2clusters}
\end{figure}

Consequently, and having checked that more than one representatives offers better results, the next decision was to perform our analysis using 2 or 3 representatives, i.e. 2 or 3 clusters. For this decision, we trust on the quality measures (silhouette~\cite{rousseeuw1987silhouettes}) to check which alternative offers better results according to the independence among clusters. Although Figure~\ref{fig:2clusters} only shows the results for Mondays divided into 30 min., the results are similar independently of the time period or the day of the week. Thus, regarding this criterion it seems that using 2 clusters supports better results.

\subsection{Selecting the sampling frequency and geographic granularity}
\label{sec:frecuency}

Recalling \figurename~\ref{fig:abnormalDayDetectionResults}, there were two additional parameters to take into account during the process: the sampling frequency of the Instagram posts, and the geographic granularity derived from the division of the space into different cells to ease the centroid location phase. 

We tried different combination of values for these two parameters: 30 minutes frequency when using a grid of 5x5 cells, 15 minutes frequency when using this same 5x5 grid (higher temporal frequency, same spatial granularity), and also 15 minutes frequency using a granularity of 7x7 grid (higher temporal frequency, and higher spatial granularity). 

Whereas for the case of one centroid, the three combinations perform very similarly, being the 5x5 grid sampled every 30 minutes the one performing slightly better, in the case of two centroids it is clear that the 15 minutes sampling frequency combined with the 7x7 grid performs better in all cases, achieving a higher percentage of detected abnormal days in a lower number of processed days. This can be due to the fact that using two different centroids is a way to divide the city into two zones. Therefore, the movement of the centroids is more fine grained, which requires a higher spatial granularity to be able to faithfully capture the differences in the movement of the centroids. Then, since a more fine grained grid is needed, it is also necessary to increase the sampling frequency in order to have a higher number of samples.

\subsection{Choosing an entropy window}
\label{sec:entropywin}

\begin{figure*}[!htbp]
  \centering
    \includegraphics[width=\textwidth]{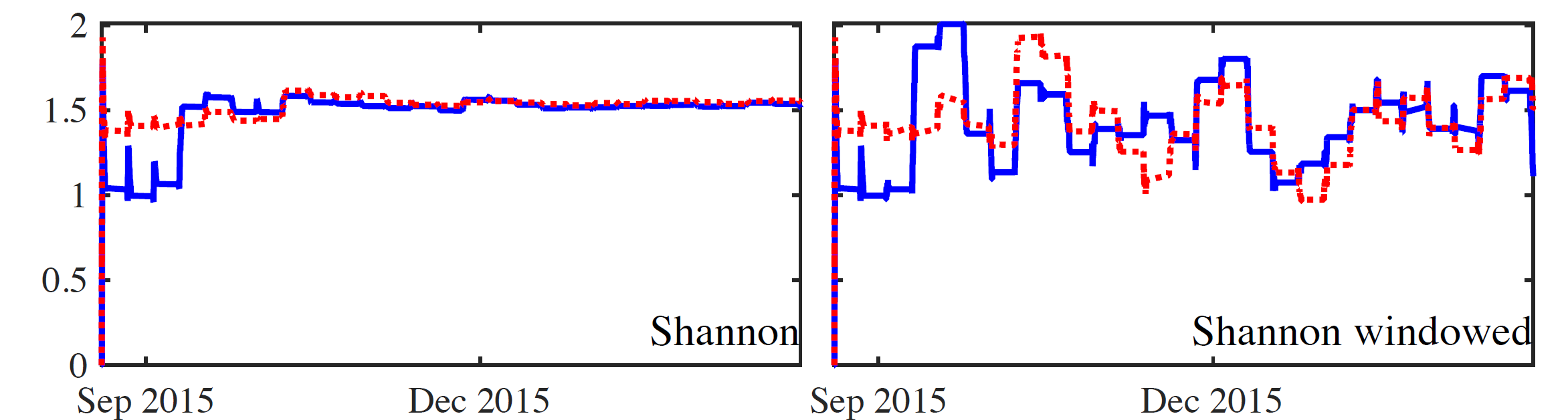}
	\caption{Use of entropy window (evolution of Thursdays,
	considering 2 centroids, a sampling frequency of 15 minutes, and
	the region split into 7x7 grid)}
\label{fig:thursdayEntropies7x715minWindows}
\end{figure*}

When calculating entropy, we realised that considering the location sequence from the beginning to each interval, $i$, led to very small
variations in the result
(Figure~\ref{fig:thursdayEntropies7x715minWindows} left). As more
samples are available to calculate $p(l,i)$ (see
equation~\ref{eqn:grassbergerEntropy2}), more samples are needed to notice a change, whereas unexpected events last, at most, one day (48 or 96 samples). Therefore, if there are two identical events happening, the first one in time will have a stronger impact in $p(l,i)$ than the last one in time. For this reason, we tested a windowed version of the entropy calculation (Figure~\ref{fig:thursdayEntropies7x715minWindows} right) with window sizes, $W$, of 4 weeks (e.g., entropy is calculated using the samples of 4 consecutive Thursdays). Different values of $W$ were tested, with best results obtained for $W = 4$.

\subsection{Selecting an entropy estimator}
\label{sec:entropyestimator}

\begin{figure*}[!htbp]
  \centering
    \includegraphics[width=0.8\textwidth]{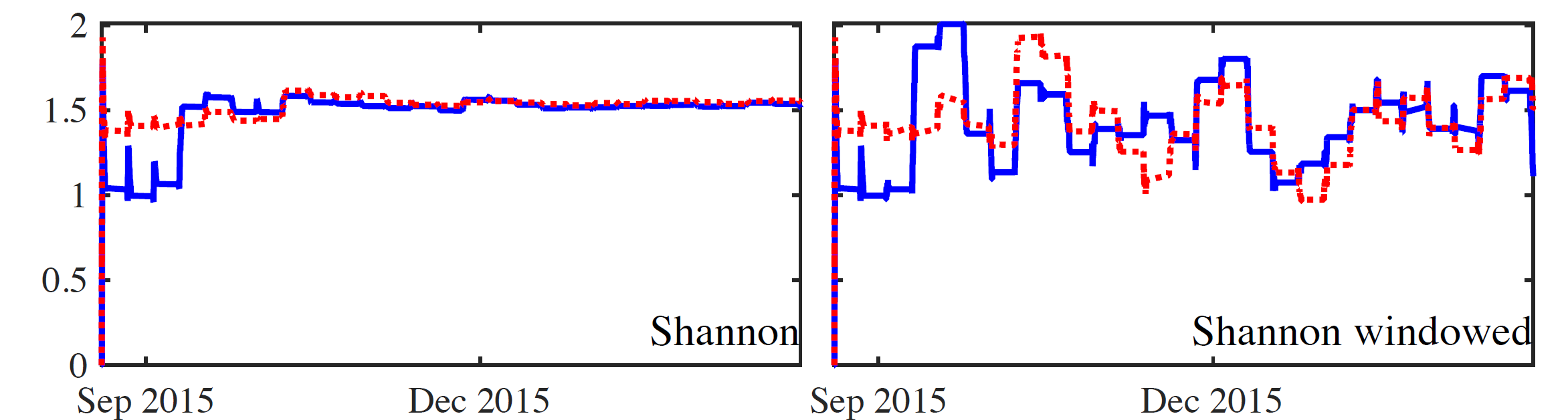}
	\caption{Entropy evolution of Thursdays, considering 2 centroids, a sampling frequency of 15 minutes, and the region split into 7x7 grid}
\label{fig:thursdayEntropies7x715min}
\end{figure*}

Finally, different estimators of the entropy can be used (see Section~\ref{sect:entropy}). We have tested the Shannon entropy introduced by Claude E. Shannon, the Hartley or maximum entropy, and the entropy rate estimator
proposed by Grassberger. The results obtained for the three estimators of the entropy, with or without windowing, are shown in Figure~\ref{fig:thursdayEntropies7x715min}. The Shannon and Grassberger estimators with a windows of 4 weeks lead to greater differences in the entropy estimation when there are days with unexpected events, so we decided to use the Shannon estimator for the rest of the analysis since it is faster to compute for each new symbol received.

\section{Results and analysis}
\label{sect:results}

According to the analysis detailed in the previous section, we have selected the most convenient parameters for the methodology: (i) an hybrid approach that combines K-means and DBSCAN to identify 2 clusters and their centroids, which will be the representative points; and (ii) the temporal and geographic granularity will be determined by slots of 15 min. and 7x7 grids (although results for 30 min. and
5x5 grids will also be presented). We have performed a two-folded
assessment where the effectiveness of the methodology is detailed
(Section~\ref{sect:efficacy}) and where efficiency is also analyzed
(\ref{sect:efficiency}).

\subsection{Effectiveness of the methodology}
\label{sect:efficacy}

Firstly, we need to define which is the entropy variation threshold between two consecutive days to consider that the change is high enough to consider the second day as unusual or abnormal. That entails analyzing the information that is represented in Figure~\ref{fig:thursdayEntropies7x715minWindows}. With this aim, we have calculated this entropy variation, difference between the entropies of any two consecutive days, for the complete data set to obtain a ranking. The top position is the day with highest entropy variation with the previous one, and the last position is the day with lowest entropy variation with the previous one.

\begin{figure*}[!htbp]
  \centering
   \begin{tabular}{cc}
   \includegraphics[width=0.5\columnwidth]{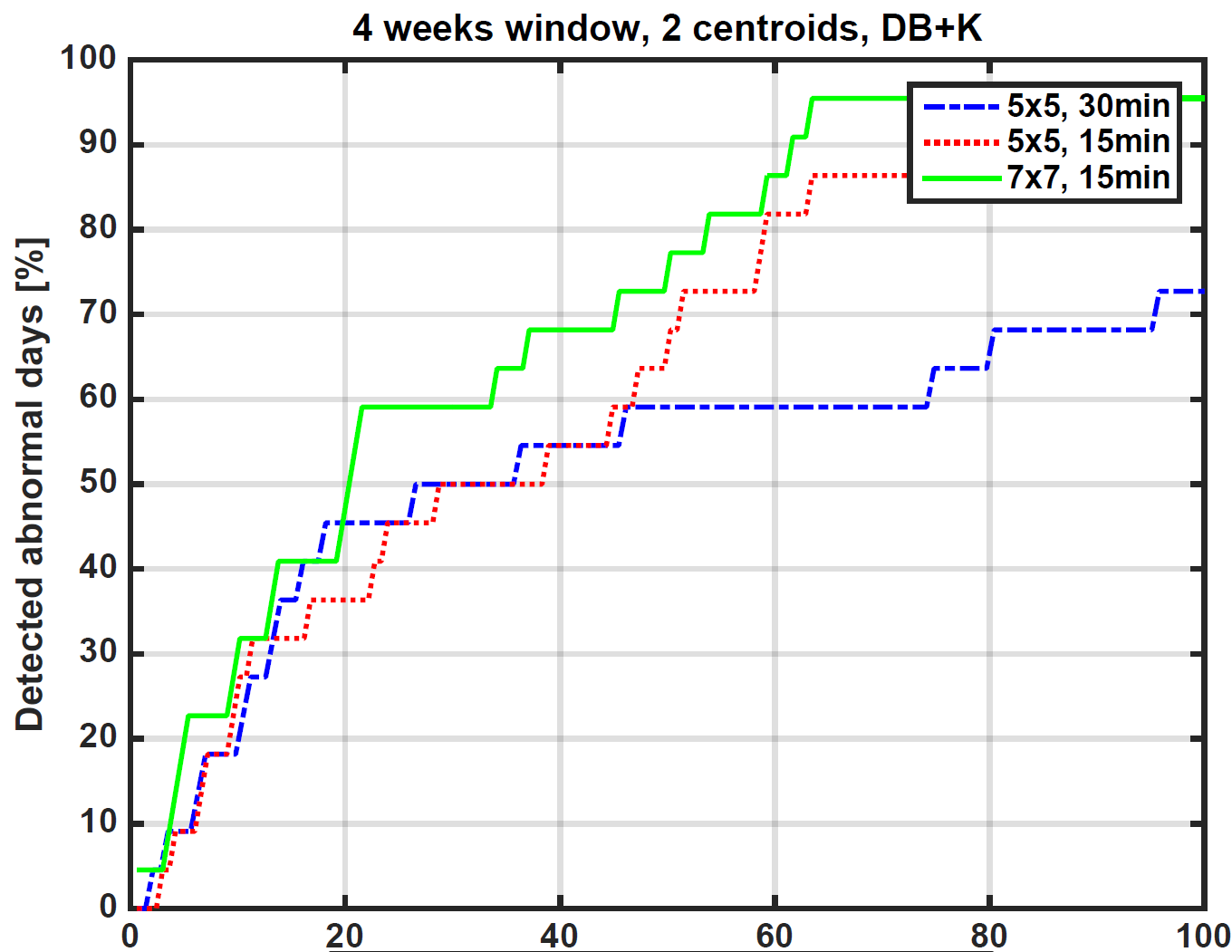}&
   \includegraphics[width=0.5\columnwidth]{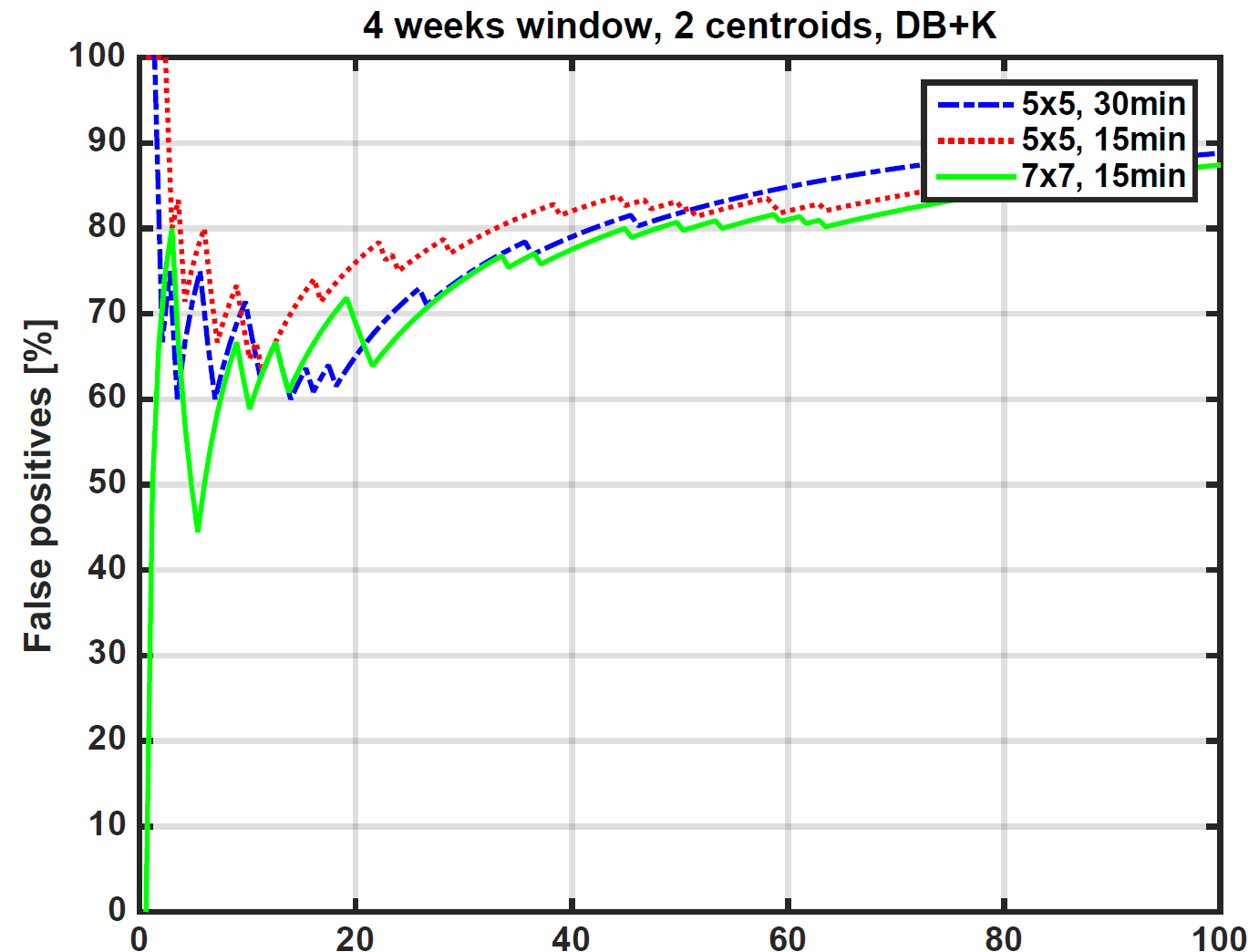}\\
   \includegraphics[width=0.5\columnwidth]{abnormalDaysShannon6weeks2C_DBK.png}&
   \includegraphics[width=0.5\columnwidth]{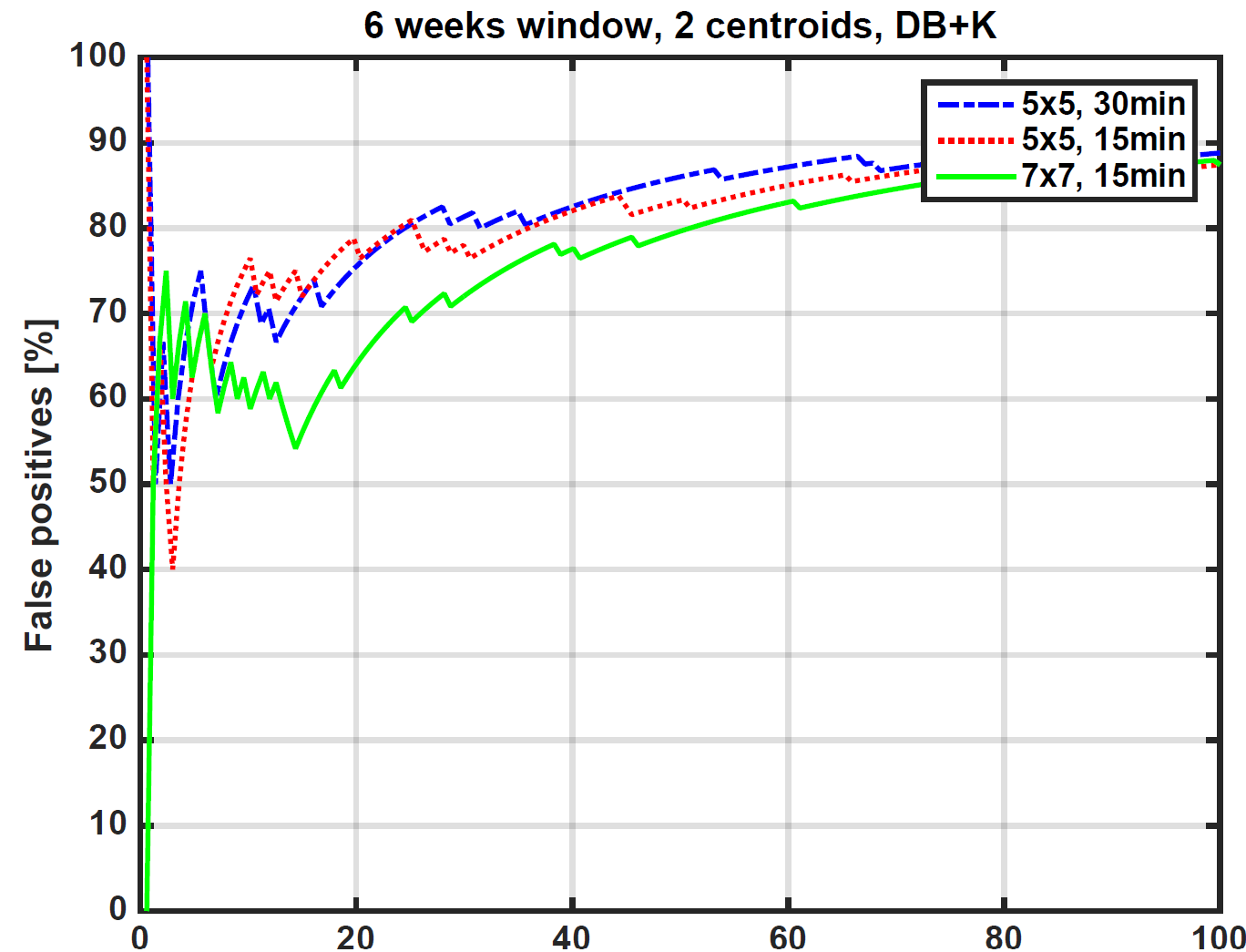}\\
   \end{tabular}
	\caption{Abnormal day detection with two centroids using
	DBSCAN+K-means a) detection rate with 4 weeks window, b) false
	positives rate with 4 weeks window, c) detection rate with 6 weeks window, d) false positives rate with 6 weeks window,}
	\label{fig:abnormalDayDetection2COriginal}
\end{figure*}

After that, we have compared the obtained ranking with the information in
Table~\ref{table:specialEvents}, i.e. the days that we have detected with abnormal behavior applying our previous research work~\cite{dominguez2017sensing}. Figure~\ref{fig:abnormalDayDetection2COriginal} depicts the results by representing the percentage of special days detected when considering the ranking of entropy variation. That is, for instance, if there were 10 special days out of 100 total days (in the table), and in the first 10 positions of our ranking there were 8 special days, that entails our new approach would detect the 80\% of the cases (y-axis) when considering 10\% (x-axis) of the total number of days in the data set. Therefore, in the ideal case that we pursue, the result should be the 100\% of special days detected when analyzing as many days as there are in the table of special days (10 events).

Consequently, Figure~\ref{fig:abnormalDayDetection2COriginal} plots the results of this analysis for different values of temporal and geographical granularity, and for different entropy windows $W$. With $W$=4 weeks we can spot up to 55\% of special days in the first 20\% of the total ordered days. Therefore, in order to identify the highest number possible by considering the least number of total days, a window of 4 weeks is preferable, combined with $T=15$ minutes and any grid size (both $S$=5x5 and $S$=7x7 overlap). Besides these results, something even more interesting came up during the analysis. Taking a look at the steepest changes in entropy (the top values in the ordered list), we further analyzed the contents of the posts and discovered that three of the days at the top of the list corresponded to an unknown event for us: the ComiCon conference, held in New York during October the 8th to the 11th. That discovery ignites the expectations regarding the method proposed as one to quickly capture unexpected behaviors in the city.

Once the two parameters (temporal interval and square size) have been established, the sequence of symbols (one sequence per day of the week) is processed to quantify the change, if any, in the crowd behavior. In order to do so, the entropy of the sequence at each step is estimated to have an idea on how the randomness of the sequence changes. If the sequence becomes suddenly highly random, that reflects an important change in the crowd behavior. The same applies if the sequence becomes suddenly very predictable. This randomness or predictability is what entropy measures.

Finally, inspecting the entropy of the sequence representing each day of the week, we extract the highest changes as abnormal days. To evaluate
the process, we use a list of holidays and other known abnormal days so as to measure the accuracy and false positives ratio (Figure~\ref{fig:abnormalDayDetection2COriginal}). However, we have no information about other possible abnormal days that could have happened, and thus the evaluation is just conservative.

\subsection{Efficiency of the methodology}
\label{sect:efficiency}

Improving the efficiency was our main objective when we started to work in a new procedure for anomalies detection, so in this section we compare both options: (i) our previous methodology~\cite{dominguez2017sensing}, based on density clustering, and (ii) the new approach introduced in this paper, based on entropy variations.

Our previous approach required of three main calculations to work. First, a stage to obtain the behavioural patterns , i.e. the reference clusters or patterns of the city. This first stage is not needed in our improved approach, at least not for detecting if something unusual is happening in the urban area. Second, our previous approach requires a DBSCAN analysis over the on-the-fly data gathered from LBSNs, whose complexity can be measured as $O(n \cdot \log n)$, with $n$ the number of geo-located points to be clustered in the time slot. Finally, it is also needed to compare both clusterings: (i) the one obtained when analyzing the data on-the-fly and (ii) the reference clustering, which represents the typical behavior for the day and time-slot analyzed. This complexity can be assessed as $O(N_d \cdot N_p \cdot n_{av}^{2})$, where $N_d$ is the number of detected clusters in the DBSCAN analysis, $N_p$ is the number of the clusters in the pattern of the city and $n_{av}$ is the mean of data points in the identified clusters.

The new approach, as mentioned above, does not need any training phase to detect unexpected behaviors in the city. It needs only the three following contributions. First, a DBSCAN clustering to detect the two most relevant clusters, whose complexity can be measured as $O(n \cdot \log n)$. Second, a K-Means clustering to obtain the two representative points, which entails a complexity of $O(n)$. Finally, the complexity of the entropy estimation. This complexity first depends on the cardinality of the alphabet $|\sX|$ (i.e., the number of possible different symbols in our distribution). In our case, the
cardinality of the alphabet is the number of cells in the grid. Thus, sssuming a grid of $L \times L$ cells, it entails that $|\sX| = L^2$. Estimating the Shannon entropy of a discrete $k$-symbol distribution requires $O(k/log k)$ samples, in our case $O(L^2 / 2 log L)$. Recently, it has been demonstrated that using a more general Rényi entropy (\cite{ACHARYA2015}), it would require just $O(L)$ samples. Therefore, and since our proposal is working with values of $L= 5$ or $L= 7$, we are facing a really low value for the initial sample sequence (transitory period). Additionally, and given a sequence of samples of length $N$, the complexity of this estimator is $O(N)$ (\cite{ACHARYA2015}). However, this is the complexity for processing a whole $N$ samples sequence (all the history of representatives) at a time. However, in our algorithm, we only need to process a new symbol (set of representatives) each interval time, so we can significantly reduce the complexity
of the algorithm using the fast entropy estimator that we have proposed in~\cite{rodriguez2015analysis}. This estimator is based on a Lempel-Ziv (LZ) prediction algorithm, where previous discovered patterns are stored in a tree data structure that makes it faster to look
for already seen patterns. Note that this algorithm increases its efficient when we do not consider an entropy window (see section~\ref{sec:entropywin}). When we define an entropy window $W$, a new sample sequence must be considered, not just adding the new symbol at the end, but also deleting the first (oldest) one, but as $W$ is constant, the complexity of estimating the entropy when a window of symbols is used is O(1).

Therefore, this new hybrid approach has less complexity, requiring less computational load to detect anomalies in the citizens behaviour. Consequently, the efficiency of our new approach is better independently on the specific hardware-software equipment used to run the algorithms.

\section{Conclusions and further work}
\label{sec:conclusion}

Users, as sensors on the move, provide huge amount of data through their mobile devices that might supplement the information gathered from the infrastructure sensors that are already installed in smart cities. In our approach, the real-time geo-located information shared by users in LBSNs is used to early detect unexpected events in urban areas. More specifically, we analyze (24h-7d) the GPS info linked to LBSNs posts to know the citizens' activity all around the area under study. Activity in LBSNs reflects the crowd dynamics in urban areas, so when the former is the expected, the latter should be also the usual one. Consequently, changes in the activity in social media are good evidences of unexpected crowd distribution in the city.

With this aim, we propose a hybrid  methodology that combines entropy analysis with clustering techniques to detect these anomalies by analyzing the entropy variation in the locations over time of a small set of city representatives. Precisely, in this paper we studied (i) how to select these city representatives from the available data; (ii) how to select the most appropriate frequency for the entropy analysis; (iii) how to select the most appropriate geographic granularity to monitor the movement of these representative points along the city and (iv) which is the best entropy estimator and entropy windows for the process. In fact, and with the data we used for our experiments ($4,335,880$ posts gathered from Instagram in New York City during 190 days), the best parameters are 2 representative points moving along a grid of 7x7 cells (the city) and being analyzed each 15 minutes. Finally, we applied the Shannon entropy and a windowed approach of four weeks, since after different comparisons this combination provided the best results.

However, this new approach is not able to give some interesting information, like the specific area in the city where the unexpected behavior is happening or the reason why this behavior is considered as abnormal (excessive activity, lack of activity, activity located in areas where it was not expected, etc.). Thus, our proposal is using the entropy-based detection methodology as a quick procedure to early detect outliers that, in this case, it should trigger a most deep analysis.

We have compared this new methodology to our previous work~\cite{dominguez2017sensing} from two different perspectives: efficiency and effectiveness. Regarding its efficiency, the new detection methodology has less complexity than the previous one. Regarding its effectiveness, the new methodology detects all the anomalies that were detected by the previous one. Additionally, it also detects new unexpected behaviors, providing a conservative approach. This is specially interesting to assure no anomalies are going to be missed in the process, by triggering a deeper analysis when needed.

We are currently working on validating the results in different types of cities and wider areas. We can also take into account other behavioral aspects of the city that have not been taken into account yet, like seasonality.  We are studying to what extent seasonality has influence in the results and how to dynamically define the size of the window used for the entropy analysis to fit these changes and provide accurate results. The mechanism proposed in this paper could also be applied to other sources of crowd mobility traces, such as the ones available in the CRAWDAD community ({\tt http://crawdad.org/}). Finally, we are also working on combining this new approach with the analysis of the text in the posts \cite{cerezo2018discovering} that would improve the information about the events that are happening in the city to discard or not a new alert.

\section*{Acknowledgement}
\noindent This work is funded by: the European Regional Development Fund (ERDF) and the Galician Regional Government under agreement for funding the Atlantic Research Center for Information and Communication Technologies (AtlantTIC), the Spanish Ministry of Economy and Competitiveness under the National Science Program (TEC2014-54335-C4-3-R, TEC2014-54335-C4-2-R, TEC2017-84197-C4-3-R and TEC2017-84197-C4-2-R), and by the Madrid Regional Government eMadrid Excellence Network (S2013/ICE-2715).


\bibliographystyle{ieeetr}

\bibliography{references-whole}


\end{document}